# Celestial calendar-paintings and culture-based digital storytelling: cross-cultural, interdisciplinary, STEM/STEAM resources for authentic astronomy education engagement

*Annette* S. Lee[1], *William* Wilson[2], *Jeff* Tibbetts[3], *Cark* Gawboy[4], *Anne* Meyer[5], *Wilfred* Buck[6], *Jim* Knutson-Kolodzne[7], *David* Pantalony[8]

[1]Department of Physics and Astronomy, St. Cloud State University, 740 Fourth Ave. S., St. Cloud, Minnesota, 56301, USA; Fond du Lac Tribal and Community College, 2102 14th Street, Cloquet, Minnesota, 55720, USA

[2]Consultant, Ojibwe language and culture; visual artist, Minnesota, USA

[3]Fond du Lac Tribal and Community College, 2102 14th Street, Cloquet, Minnesota, 55720, USA

[4] (Emeritus) College of St. Scholastica, Department of Am. Ind. Studies, 1200 Kenwood, Duluth, MN, 55811, USA

[5] Consultant, visual artist, art educator Minnesota, USA

[6]Manitoba First Nations Education Resource Centre, 2-1100 Waverly St., Winnipeg, Manitoba R3T 3X9, Canada

[7]Director of the SCSU American Indian Center, St. Cloud State University, 740 Fourth Ave. S., St. Cloud, Minnesota, 56301, USA

[8]Curator, Physical Sciences and Medicine, Conservateur, Sciences physiques et médecine Ingenium - Canada's Museums of Science and Innovation, Musées des sciences et de l'innovation du Canada, Ottawa, Ontario

**Abstract:** In D(L)akota star knowledge, the Sun is known as *Wi* and the Moon is *Han-Wi*. They have an important relationship, husband and wife. The pattern of their ever-changing relationship is mirrored in the motions of Sun and Moon as seen from our backyards, also called the lunar phases. The framework of the cultural teaching is storytelling and relationships. Cultural perspectives in astronomy such as this remind us of how indigenous ways of knowing are rooted in inclusion, engagement, and relevancy. Designed by A. Lee in 2007, the *Native Skywatchers* initiative seeks to remember and revitalize indigenous star and earth knowledge, promoting the native voice as the lead voice. The overarching goal of *Native Skywatchers* is to communicate the knowledge that indigenous people traditionally practiced a sustainable way of living and sustainable engineering through a living and participatory relationship with the above and below, sky and earth. In 2012 two indigenous star maps were created: the *Ojibwe Giizhig Anung Masinaaigan*-Ojibwe Sky Star Map (A. Lee, W. Wilson, C. Gawboy), and the D(L)akota star map, *Makoce Wicanhpi Wowapi* (A. Lee, J. Rock). In 2016, a collaboration with W. Buck of the Manitoba First Nations Resource Centre (MFNRC), produced a third star map: *Ininew Achakos Masinikan*-Cree Star Map Book. We aim to improve current inequities in education for native young people especially through STEM engagement, to inspire increased cultural pride, and promote community wellness. Presented here will be recently created resources such as: astronomical calendar-paintings and short videos that exist at the intersection of art-science-culture. As we look for sustainable ways to widen participation in STEM, particularly in astronomy education, part of the conversation needs to consider the place for art and culture in STEM.

# 1 Introduction

## 1.1 Overview and Need

Despite enormous effort and millions of funding dollars to improve K-12 STEM education, inequities persist across the US. Nowhere is the inequity more pronounced than with Native American students. Native students are more likely to drop out of high school and more likely to be placed in special education than any other race nationally (Allen, 1997). There are over 20,000 native students in Minnesota, (2% of the total students), yet only 249 native teachers are working in the K-12 schools (0.4%) (MDE 2017). Educational inequity is evident in high school graduation rates. For years, Minnesota has been near the bottom of the list for on-time graduation rates of students of color. Fewer than 50% of native high school students graduate on time (Post, 2015), while 85% of white students in Minnesota graduate on time. Minnesota has one of the nation's greatest educational disparities between majority students and students of color (Post, 2015).

In 2009 the Minnesota Department of Education released the official State Science Standards for K-12 education. One standard requires teachers to include how underrepresented cultural groups have contributed to science, "Men and women throughout the history of all cultures, including Minnesota American Indian tribes and communities, have been involved in engineering design and scientific inquiry." (MDE 2009) Benchmark 3.1.3.2.1 specifically states "Understand that everybody can use evidence to learn about the natural world, identify patterns in nature, and develop tools. For example: Ojibwe and Dakota knowledge and use of patterns in the stars to predict and plan." (MDE 2009)

In addition to state science standards that promote cultural relevancy, each school district in Minnesota that has ten or more American Indian students can apply for Federal and State funding through the MN Office of Indian Education. Currently the Office administers the American Indian Education Aid program, which serves 143 eligible school districts, charters, and Bureau of Indian Education tribal contract schools. The number of school districts that do apply for the grants has increased every year. Each school district's superintendent and American Indian Advisory Parent Committee decide how to use the funding. Some hire staff to operate an American Indian Education program, others assign a district employee to oversee the grants. There is a loose network of American Indian education coordinators, directors, advocates, or district staff that meet on their own. The MDE Indian Education department provides moral support, education materials resources and advice. Usually it is up to the American Indian parents in a school district to put pressure on the school district to meet the needs of their children. The *Native Skywatchers* research and programming initiative works closely with many of the American Indian educator liaisons throughout the state delivering educator training workshops and indigenous STEM resources.

## 2 *Native Skywatchers* Goals and Objectives

### 2.1 Resource Development and Delivery

In 2007 A. Lee created the *Native Skywatchers* (NSW) research and programming initiative with the vision of revitalizing the Ojibwe and D(L)akota native star knowledge. The overarching goal of NSW is to communicate the knowledge that indigenous people traditionally practiced a sustainable way of living and sustainable engineering through a living and participatory relationship with the above and below, sky and earth. The aim of NSW is to improve current inequities in education for native young people, to inspire increased cultural pride, and promote community wellness. The hope is to rekindle or deepen a sense of awe and personal relationship to the cosmos for all people.

> *"I used to look up and see the Greek constellations, like the Big Dipper, or Leo the lion…but now I know that there are stars up there that are ours. It does something to me inside, to have that relationship with the stars. It's like finding a long-lost relative." - J. Tibbetts, 2010*

Funding awarded through NASA – MN Space Grant (MNSG), St. Cloud State University (SCSU) and Fond du Lac Tribal and Community College (FDTLCC) from 2009-2015 allowed NSW to develop resources and deliver community based programming over a six-year period. In 2012, for the premier *Native Skywatchers* (NSW) Educator & Community Workshop, two native star maps were designed and delivered: 1.) *Ojibwe Giizhig Anung Masinnagan* (Fig. 1) (A. Lee, et al., 2012), and 2.) *D(L)akota Makoce Wicanhpi Wowapi* (Fig.2) (A. Lee, et. al, 2013).

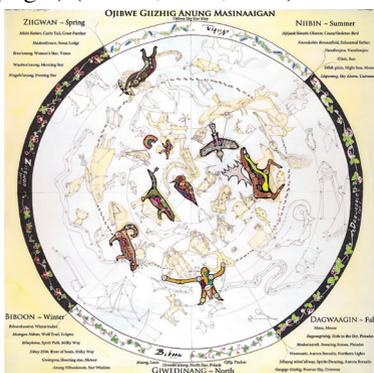

**Fig. 1** *Ojibwe Giizig Anung Masinaaigan,* A. Lee et. al., 2012

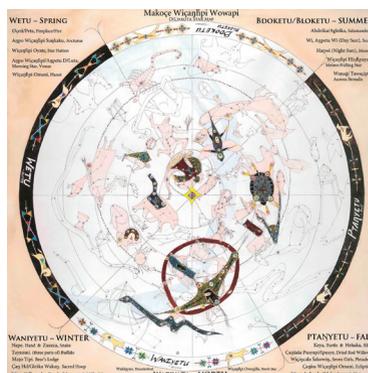

**Fig. 2** *D(L)akota Makoce Wicanhpi Wowapi,* A. Lee, et. al., 2012

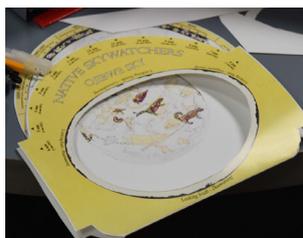

**Fig. 3**- NSW Ojibwe Planisphere

Using the native star maps as a foundational resource, many additional related resources were designed and delivered to educators and communities throughout the state, examples include: constellation guidebooks, make-it-yourself planispheres, lesson plans and worksheets compiled in a *Native Skywatchers* Curriculum Workbook, audio recording of native celestial vocabulary, power point presentations, etc. (Figs. 3-7).

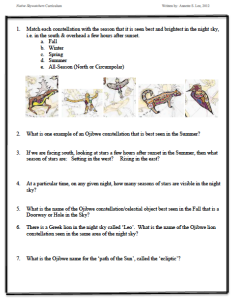 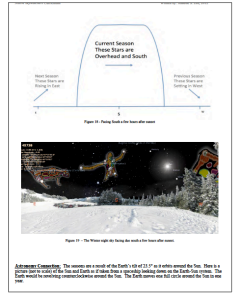 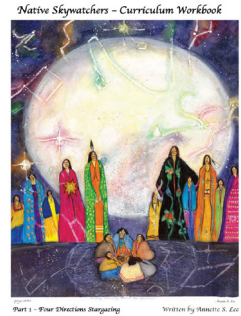

**Figs. 4 & 5** - Sample NSW Worksheets          **Fig. 6**-NSW Educator Workbook Cover

### 2.2 *Native Skywatchers (*NSW) Expansion Efforts

The NSW approach explores similarities between native ways of knowing the natural world and those of western science, toward the co-creation of a "dual-learning" system. Guided by the teaching "As it is above; it is below," the Earth-Sky relationship serves as the foundational framework of the NSW approach (Fig. 7). In this framework, the sky and stars are located above, the Earth (ground) is located below, and *Anishinaabeg*—the People—are fixed firmly and equally between the two, connecting them. The people are the stewards at the center of Earth and sky.

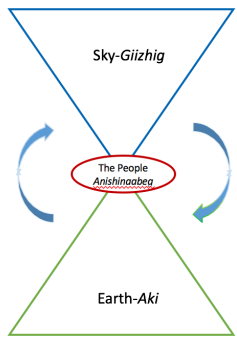

**Fig 7** Earth Sky Mirroring; *Kapemni*

*Native astronomies helped to make sense of life and relationships and reaffirm the belief in the interrelationship and interdependence of all things in an animate and living universe. Like the Earth, plants, and animals, celestial bodies are traditionally viewed by Native cultures as living beings with a creative life force that relates to and affects human beings physically and spiritually.* (Cajete, 2000)

Where the original aim of NSW was to focus on astronomical knowledge, it quickly became apparent that embedded in the sky was the reflection of the below-the earth- and the human part in both. In 2014 the original premise of the astronomy focus was expanded to encompass earth-science content and cultural knowledge, for example: water, earth, air, and fire…and/or…. hydrology, geology, atmospheric science, and energy resources.

Since 2014 NSW has received state funding through the Minnesota State Arts Board (MSAB) to design and deliver art-making interdisciplinary community based programming. Deliverables included: hands-on workshops and a touring exhibit. Across the state over 40 art-focused NSW workshops have been delivered primarily in native communities. Participants ranged in age from youth to elders. Venues included community centers, history/culture museums, schools, parks, libraries, art galleries, and medical clinics. In addition, a one-year traveling exhibit (2017-2018) showcases this work and widens the participatory circle. Over thirty, mostly native, artist-participants from professionals to beginners share their stories of a personal connection to earth and sky in this state-wide multi-media exhibit. The strength of this work is its interdisciplinary nature. Science, culture, and art, are woven together with diligent purposefulness.

   - "*The teacher (Annette) presented science as though it is connected to us in our everyday lives.*"

*-"The art we created seemed philosophical - as though the science of the skies
had a connection to how we think and feel about the world. And our art reflects
that feeling.
- "Thank you for sharing the science of our world with the native journey and
taking me in a new exciting direction."*
            *- Selected quotes, MSAB-Arts Learning participants 2015-2016*

In 2015, NSW team-members A. Lee and W. Wilson were given tobacco by a representative, Wilfred Buck, from the Ininew (Cree) nation based in Winnipeg at the Manitoba First Nations Education Resource Center (MFNERC). This collaboration yielded a third native sky map, *Ininew Achakos Masinkan*, created in 2016 and based on 30 years of community based research by W. Buck (Fig.8).

Together, A. Lee and W. Buck were invited by Canada's Science and Technology Museum to co-curate (along with permanent curator D. Pantalony) an indigenous astronomy exhibit built on the grass-roots community-based research and programming that has occurred over the past three decades. This exhibit is called *"One Sky – Many Astronomies"* and is part of the complete renovation of the museum opening Nov. 2017.

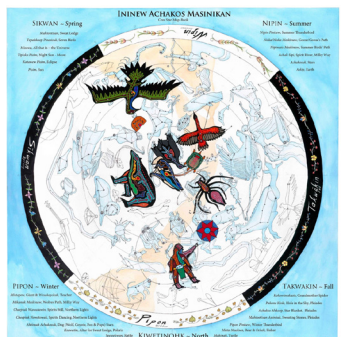

*"Through thousands of years of observation and
recognition of patterns in plants, animals, seasons and
the sky, First Peoples built knowledge about the world
around them…The sky maps and illustrations on this
wall represent stories about stars and constellations that
are deeply woven into the identity of Ininew, Ojibwe and
D(L)akota peoples."…Without reconnecting to spirit,
Achak, we risk losing the knowledge that is inherent in
Misewa, "All that is" and the idea of creation and
learning without limits, and that is never ending.*

**Fig 8** - *Ininew Achakos Masinikan*,
A. Lee, W. Wilson, W. Buck, 2016

*– W. Buck 2017, One Sky-Many Astronomies exhibit*

## 3 Native Skywatchers Approach

### 3.1 Focus on Cultural Relevancy

The NSW project draws strength from the growing body of evidence that culturally-relevant pedagogy increases student success. One recent study in Hawai'i, which gathered data from nearly 3000 students, 600 teachers and over 2000 parents, concluded:

*First, culture-based education (CBE) positively impacts student social-
emotional well-being (e.g., identity, self-efficacy, social relationships). Second,
enhanced socio-emotional well-being, in turn, positively affects math and
reading test scores. Third, CBE is positively related to math and reading test
scores for all students, and particularly for those with low socio-emotional
development, most notably when supported by overall CBE use within the
school. (Kana'iaupuni, Ledward, and Jensen 2010)*

Culturally-relevant pedagogy is grounded in the idea that culture is always present. Norms, values, beliefs, and assumptions in Western scientific and indigenous culture are welcomed and honored in a dual-learning system. A dynamic is introduced and upheld that begins with the recognition that indigenous cultural knowledge is valuable, relevant, and meaningful. Educators and students learn to recognize, utilize, and value different approaches to teaching and learning.

One example of culturally relevant pedagogy is the NSW interdisciplinary workshop, "Carving your Animal Clan", led by *Native Skywatchers* team-member J. Tibbetts (Fig. 9-12). These workshops allowed participants to learn about the Ojibwe clan system, and to learn the geologic story of the 'rocks under our feet'. Many participants were surprised that 1 billion years ago North America attempted a mid-continental split and Minnesota would have become the new 'west coast' of the eastern half of North America. Important culturally relevant teachings were layered into the discussion such as, the use of basalt in the sweat lodge ceremony *madoodiswan,* and the Ojibwe *Madoodiswan*, Sweat Lodge constellation (Corona) in the night sky. The workshops are designed to give participants the opportunity to digest the ideas of the artwork, the culture and the science and then to 'make it their own'. Participants are given the materials to carve their own animal clan. Realistically not all native or mixed-race native participants are knowledgeable about their family clan. Much tradition has been lost. Participants are encouraged to talk to local elders and to focus in the short term on an animal that has meaning to them.

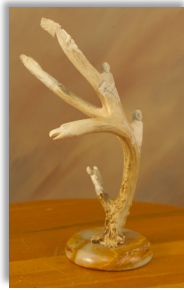

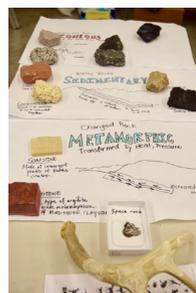

**Fig. 9** *Antler carving, J. Tibbetts*

**Fig. 10** *NSW workshop*

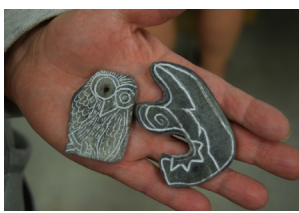

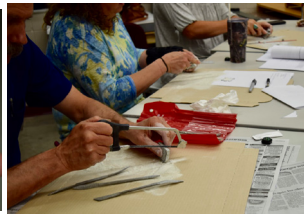

**Fig 11**-*NSW Participant carvings, 2017*

**Fig 12**- *NSW Participants, 2017*

### 3.2 Focus on Storytelling

Humans beings have relied on storytelling as a primary means of communication since our infancy as a species over 30,000 years ago. Native cultures have practiced oral tradition for tens-of-thousands of years. Medical research has shown that our brains are wired to learn and remember better by stories (Schwertly 2014). "*Research shows our brains are not hard-wired to understand logic or retain facts for very long. Our brains are wired to understand and retain stories.*" (Gottschall 2013) Even more interestingly, while the brain listens to a story, "*the brain doesn't look like a spectator, it looks more like a participant in the action.*" (Gottschall 2013) The research and community programming of the *Native Skywatchers* initiative is focused on storytelling as an effective method of learning and teaching. For example, the *Ojibwe Giizhig Anung Masinnaaigan*, native star map is an astronomical map of the night sky, a teaching tool. Woven into this map are countless stories about each of the constellations, the planets, the seasons, and the language. There are layers of stories and volumes of knowledge. There are cultural stories and personal stories. In the northern hemisphere winter night sky, a viewer wearing an Ojibwe cultural lens can see

*Biboonkeonini*-the Wintermaker. A viewer from the Western science perspective can see Orion. The Ojibwe stories in this case relate to how *"Each season has certain spirits that make that season happen."* (A. Lee et. al. 2014) and how the Wintermaker was defeated by the duck. (Gawboy 2014). Western science stories tell us of the star formation and planetary system accretion happening in the Orion Nebula. Both the Ojibwe perspective and the Western science perspective have a valuable and interesting story to tell. Here the intent is to weave both stories together in a tapestry that is dense with knowledge. Since 2016 NSW resource development includes short story-telling videos, such as *Wintermaker* and *Ojiig*, two Ojibwe constellations in the night sky, stories told by C. Gawboy, videos produced by A. Lee (Figs. 13 & 14).

**3.3 Focus Place Based Relevancy**

The NSW project emphasizes culturally-relevant place-based learning and teaching. Place-based education engages students and communities, breaks down Native students' "boredom" (Platero, 1986; Deyhle, 1989), and demonstrates that the application of science and its results have tangible local significance. *"In the natural sciences, place-based pedagogy is advocated as a way to improve engagement and retention of students, particularly members of indigenous or historically inhabited communities"* (Semken 2008). Because the NSW project focuses on culturally-relevant pedagogy and place-based research, the project roots itself in history and location while making global connections.

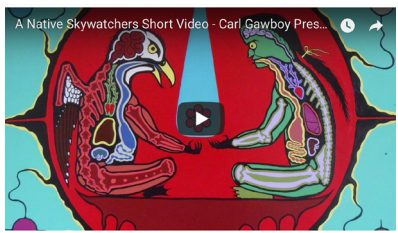

**Fig 13**- *Wintermaker Video, C. Gawyboy & A. Lee, 2016*

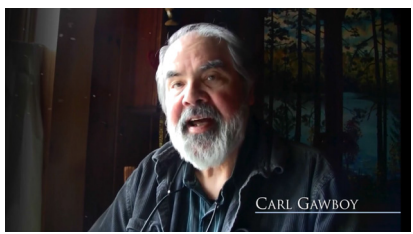

**Fig 14**- *Ojiig Video, C. Gawboy & A. Lee, 2016*

One NSW interdisciplinary workshop called, *"Be Humble for you are made of Earth"*, led by A. Meyer tells the story of the clay (Figs. 15-18). Here the science story is presented to participants. This includes the last glacial melt approximately 15,000 years ago and the formation of the 10,000 lakes that Minnesota is known for, as well as till, moraines, and Glacial Lake Agassiz. Alongside this idea is the exploration of pottery shards from the indigenous people living in the Minnesota area 3000 years ago, and some traditional Ojibwe and D(L)akota views of clay. Such as the use of catlinite/claystone (a mass of limestone found in a clay deposit, commonly called 'pipestone') to make the sacred pipe (https://en.wikipedia.org/wiki/Catlinite). In the D(L)akota star map, the constellation

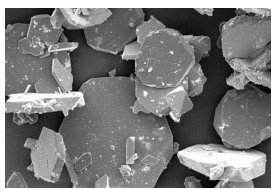

**Fig 15**-*Flat platelets of clay*

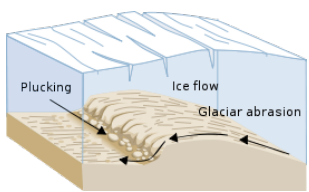

**Fig 16**-*Glacial plucking, L. Benitez, 2004*

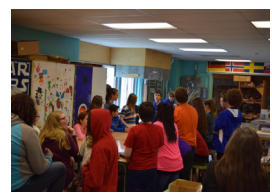

**Fig. 17**-*NSW Interdisciplinary workshop*

*Cansasa Pusyapi*, Red Willow, relates to the Pipe Ceremony. It is said, that "the bowl of the pipe comes from the blood of the people". (O'Rourke 1993)

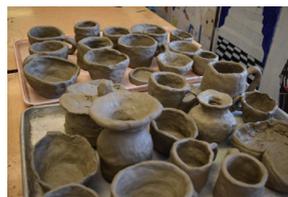

**Fig. 18**-*Participant clay works pre-firing*

### 3.4 Focus on the Visual Language

Learning to think and communicate visually has become increasingly important in both mainstream society and in academia (CSTS 2006). The scientific method itself was inspired by evidence based entirely on the visual, for example: Brahe's almanacs (Brahe 1588, 1572), and Galileo's sketches (Galileo 1610, 1632). It has been shown that scientists have an "*almost obsessive preoccupation with visual representations…*" and that "*…visual representations are central to the work creating science and to communicating science to others*." (McGinn, Roth, 1999) Another group found that: *"…children with developed visual perception, visual thought and representational skills are actually better with numbers and physical concepts"* (Stavridou & Kakana 2008). A National Academy of Sciences report determined that spatial thinking is 1.) woven into the fabric of our daily lives; 2.) integral to successful problem solving, and 3.) is "under-recognized, undervalued, underappreciated, and therefore, under-instructed." (CSTS 2006).

In 2016 the *Native Skywatchers* initiative increased and expanded efforts in the development of educational indigenous STEM resources. Amplified efforts were directed towards the Ojibwe and D(L)akota cultural perspectives of the Moon and the Sun. This request came from many K-12 teachers and/or American Indian liaisons who were already implementing the native star maps in their classrooms. In comparison to the constellation material which appears in two substrands (in the third and eight grade curriculum standards), the Sun and the Moon content focus appears in eight substrands.

Native Skywatchers (NSW) celestial calendars are a 'first step' in rethinking the established teaching and learning approach to Sun and Moon. The first Moon calendar, "Our Moon" was produced in 2016 uses a rectangular grid layout, and relies heavily on color and texture to convey the phases of the Moon for each day of the year (Fig. 19). New Moons are coal black with green shadows. Full Moons are a heavily textured off-white bathed in a crimson red wash. In 2016 a second Moon calendar was created (Fig. 20). Here the layout of a circle was chosen to more appropriately match the cyclical motion of the Moon. Imagine standing at the center of the larger circle, then looking outward radially. The phase of the

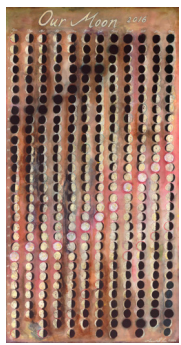

**Fig. 19**-*Our Moon, A. Lee, 2016*

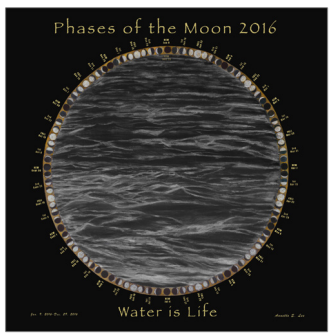

**Fig. 20**-*Water is Life Moon, A. Lee, 2016*

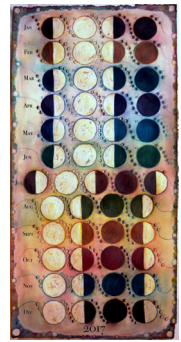

**Fig. 21**-*Our Moon, A. Lee 2017*

Moon can be seen every three days from Jan. 1, 2016 to Dec. 31, 2016. The phases move around in a clockwise direction. This celestial lunar calendar is a visual springboard for the interwoven teachings between the Moon, water, and life. In 2017 a third Moon calendar was produced with a focus on the four main Moon phases- new, first, full, third (Fig. 21). It is important that the work is relevant, and most viewers of the sky have a basic knowledge of at least one or two of the four main lunar phases. This common ground is key to building the next layers of understanding and discussion. Following the lunar calendar paintings, lesson plans, worksheets, and a short video on the Ojibwe Moons were created (Fig. 22).

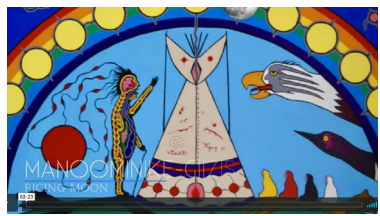

*Fig. 22 Ojibwe Moons, A. Lee & W. Wilson, 2017*

In 2016 and 2017 Sun calendars were created as an interdisciplinary learning tool (Figs. 23 & 24). Much like the Lunar Phase celestial calendars, the Sun's path for a viewer looking towards the south meridian is shown for the year. Here the local maximum and minimum hours of daylight can easily be seen to correspond the longer/highest and shortest/lowest paths of the Sun. A viewer in central Minnesota can expect about 15. 5 hours of daylight around the Summer Solstice ~June 21, and about 8.5 hrs. of daylight around the Winter Solstice ~ Dec. 21 each year.

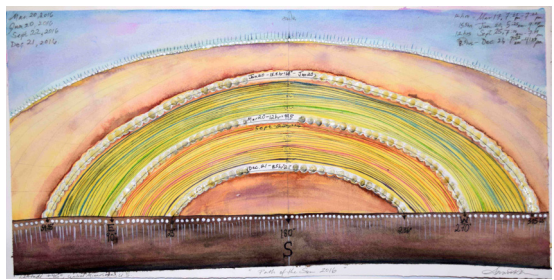

*Fig 23-Path of the Sun, A. Lee, 2016*

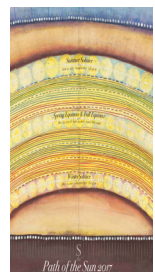

*Fig 24- Path of the Sun, A. Lee, 2017*

# 4 Conclusions

## 4.1 Widening Participation in STEM

Based on available data, Native Americans are facing a crisis in STEM education. Traditionally, Native American people saw themselves as stewards of the natural world, and there is clearly a disconnect in education systems that have failed to tap into what should be an area of interest. Minnesota Compass, a social indicators project, gleans the following from the National Assessment of Educational Progress: In 2015, Native American 4th graders reported the least interest in science of all student populations in Minnesota. In 2016, less than 40% of Native American 5th graders achieved Minnesota science standards. In 2017, Native American 8th graders ranked lowest (less than 30%) in meeting Minnesota math standards.

**4.2 Relevancy, Identity, and Cultural pride**

The NSW project fosters a dual-learning environment in which the stories told by science and native cultures (particularly Ojibwe and D(L)akota) are woven together into a tapestry—an image that ultimately reveals new and deeper meaning than could not be found in any of the threads alone. In this way, Native American students are encouraged to approach, pursue, and contribute to STEM without compromising their cultural identity. The work of the *Native Skywatchers* research and programming initiative aims to undermine this crisis. Since 2007 interdisciplinary resources that stand at the nexus of science, culture, and art have been designed and delivered by a team of native scientists, educators, cultural experts, and artists. The intent of this work is to promote the native voice as the lead voice in defining indigenous STEM. And at the same time the inclusion of culture should not diminish the science content, but rather open the door for wider participate and a fuller spectrum of teaching and learning engagement.